\documentclass[12pt]{article}

\catcode`\@=11

\global\arraycolsep=2pt
\usepackage{amsbsy,amssymb,latexsym,amsfonts,amsmath}
\usepackage{graphicx,color}

\allowdisplaybreaks

\begin{document}
\begin{flushright}
\parbox{4.2cm}
{RUP-23-4}
\end{flushright}

\vspace*{0.7cm}

\begin{center}
{ \Large Ambitwistor string amplitudes in light-like linear dilaton background}
\vspace*{1.5cm}\\
{Mayu Kato and Yu Nakayama}
\end{center}
\vspace*{1.0cm}
\begin{center}

Department of Physics, Rikkyo University, Toshima, Tokyo 171-8501, Japan

\vspace{3.8cm}
\end{center}

\begin{abstract}
 Using the ambitwistor string theory, we study graviton scattering amplitudes in a light-like linear dilaton background of ten-dimensional supergravity. At the tree level, we find that the three-graviton amplitude coincides with the type II superstring theory, and the four or more graviton amplitudes differ from the superstring theory but satisfy the scattering equations. Due to a modified momentum conservation law different at each order in genus expansion, a non-zero amplitude is determined solely from the particular order in perturbation theory without further corrections. 
\end{abstract}

\thispagestyle{empty} 

\setcounter{page}{0}

\newpage

\section{Introduction}
In quantum field theories, scattering amplitudes are usually computed as an on-shell limit of correlation functions. It turns out, however, that the scattering amplitudes sometimes show more elaborate structures that are obscured in off-shell correlation functions. For instance, graviton  scattering amplitudes in supergravity satisfy the so-called scattering equation \cite{Cachazo:2013iaa}\cite{Cachazo:2013gna}, which suggests the origin of the underlying string-like theory, where the scattering amplitudes can be computed directly without referring to off-shell quantities.

The ambitwistor string theory gives the worldsheet interpretation of the scattering equation in supergravity \cite{Mason:2013sva}\cite{Berkovits:2013xba}. The ambitwistor string theory is defined on a two-dimensional worldsheet but it is not a conventional string theory. It is supposed to give scattering amplitudes in supergravity without massive modes.
The natural question to be asked is if such structures remain valid in non-trivial supergravity background beyond the flat Minkowski space-time. In particular, when the non-trivial background is a solution of the superstring theory, we may be able to see the difference between the superstring theory and ambitwistor string theory by comparing the scattering amplitudes e.g. in the black hole background or the cosmological evolution, where the quantum gravity seems to play a significant role.
Some promising results have been obtained in \cite{Adamo:2014wea}\cite{Adamo:2017sze}\cite{Adamo:2018ege}\cite{Roehrig:2020kck}\cite{Eberhardt:2020ewh}. In this paper, we study yet another example of scattering amplitudes in the non-trivial supergravity background from the ambitwistor string theory. 

The background we will study in this paper is a light-like linear dilaton background \cite{Ho:2007ar}\cite{Chan:2009eh}.
The light-like linear dilaton background has attracted some theoretical interest. It is an exact string background beyond supergravity, albeit it has a strong coupling region where the perturbative string computation may not be trusted. It is a time-dependent background, so we 
 may be able to learn the nature of string theory in the time-dependent background, which is typically very difficult. It may have potential implications for string cosmology or the beginning of the universe. To investigate these aspects, the Matrix theory descriptions and high-energy behaviors of the string theory in the light-like linear dilaton background have been pursued in \cite{Craps:2005wd}\cite{Craps:2006xq}\cite{Li:2005sz}\cite{Li:2005ti}\cite{Chen:2005mga}\cite{Chan:2009kb}. The comparison with the ambitwistor string theory should enable us to understand the fundamental difference between superstring theory and supergravity in the time-dependent background.

The organization of the paper is as follows. In section two, we review the abmitwistor string theory in a curved background. In section three, we introduce a light-like linear dilaton background and compute graviton scattering amplitudes in the ambitwistor string theory. In section four, we give further discussions and a conclusion.

\section{Ambitwistor string theory in curved background}
The ambitwistor string theory in the curved background is described by the worldsheet theory
\begin{align}
 S = \frac{1}{2\pi} \int d^2z \left(P_\mu \bar{\partial} X^\mu + \bar{\psi}_\mu \bar{D} \psi^\mu \right) \ , 
\end{align}
together with  {\it two} copies of left-moving ghost $(b,\tilde{b},c,\tilde{c})$ and superghost $(\beta,\bar{\beta},\gamma,\bar{\gamma})$ {\it without right-moving modes}. Here the covariant derivative $\bar{D} \psi^\mu = \bar{\partial} \psi^\mu + \Gamma^{\mu}_{\ \nu \rho} \psi^\nu \bar{\partial} X^\rho$ is with respect to the Levi-Civita connection $\Gamma^{\mu}_{\ \nu \rho}$ of the target space-time. It is more convenient to write the action in a form similar to the curved-space beta-gamma system:
\begin{align}
S = \frac{1}{2\pi} \int d^2z \left(\Pi_\mu \bar{\partial} X^\mu + \bar{\psi}_\mu \bar{\partial} \psi^\mu \right) \ , 
\end{align}
where $\Pi_\mu = P_\mu + \Gamma^{\lambda}_{\mu \nu} \bar{\psi}_\lambda \psi^\nu$. We have assumed that the worldsheet is (locally) flat, but we will later introduce the analogue of the Fradkin-Tseytlin term (i.e. curvature coupling to the background dilaton) when needed. 

Then the worldsheet operator product expansion can be computed as
\begin{align}
X^\mu(z_1) \Pi_\nu(z_2) &\sim \frac{\delta^\mu_\nu}{z_{12}} \cr 
\psi^\mu(z_1) \bar{\psi}_\nu(z_2) &\sim \frac{\delta^\mu_\nu}{z_{12}} \ , 
\end{align}
where we have defined $z_{12} = z_1 - z_2$. The other operator product expansions, most importantly $X^\mu(z_1) X^\nu(z_2)$, are non-singular. 
With these variables, there is no reference to the target space metric unlike in the worldsheet formulation of the superstring theory. It will appear in the gauging currents that we introduce now.

In the BRST quantization of the ambitwistor string theory, the matter current $T$,$\mathcal{G}$,$\bar{\mathcal{G}}$ and $\mathcal{H}$ are effectively gauged. In the background where the Kalb-Ramond field vanishes, they are given by (see \cite{Adamo:2018ege} for the detailed account)
\begin{align}
T =& \Pi_\mu \partial X^\mu -\frac{1}{2} (\bar{\psi}_\mu \partial \psi^\mu + \psi^\mu \partial \bar{\psi}_\mu) - \frac{1}{2} \partial^2 \log (e^{-2\Phi} \sqrt{g}) \cr
\mathcal{G} =& \psi^\mu \Pi_\mu + \partial (\psi^\mu \Gamma^{\kappa}_{\ \mu \kappa} ) - 2 \partial (\psi^\mu \partial_\mu \Phi) \cr
\bar{\mathcal{G}} = &g^{\mu\nu} \bar{\psi}_\nu (\Pi_\mu - \Gamma^{\kappa}_{\ \mu \lambda} \bar{\psi}_\kappa \psi^\lambda) -g^{\mu\nu} \partial( \bar{\psi}_{\kappa} \Gamma^{\kappa}_{\ \mu \nu}) - 2 \partial(g^{\mu\nu} \bar{\psi}_\mu \partial_\nu \Phi) \cr
\mathcal{H}  =& g^{\mu\nu}(\Pi_\mu - \Gamma^{\kappa}_{\ \mu \lambda} \bar{\psi}_\kappa \psi^\lambda)(\Pi_\nu - \Gamma^{\kappa}_{\ \nu \lambda} \bar{\psi}_\kappa \psi^\lambda) -\frac{1}{2} R^{\kappa \lambda}_{\ \ \mu \nu} \bar{\psi}_\kappa \bar{\psi}_{\lambda} \psi^\mu \psi^\nu \cr 
&-g^{\mu\nu} \partial(\Pi_\rho \Gamma^{\rho}_{\ \mu \nu} - \bar{\psi}_\kappa \partial \psi^\lambda g^{\mu\nu} \partial_\lambda \Gamma^{\kappa}_{\ \mu\nu} + \psi^{\mu} \partial_\mu (g^{\rho\sigma} \partial (\bar{\psi}_{\kappa} \Gamma^{\kappa}_{\ \rho \sigma})) \cr
&-2\partial(g^{\mu\nu} \Pi_\mu \partial_\nu \Phi) - \partial(\bar{\psi}_{\kappa} \psi^\lambda(2\nabla^{\kappa}\partial_{\lambda} \Phi - 2g^{\mu\nu} \Gamma^{\kappa}_{\ \mu \lambda} \partial_\nu \Phi)) \ . \label{currents}
\end{align}
Here, $g_{\mu\nu}$ is the target space-metric and $R^{\mu\nu}_{\ \ \rho\sigma}$ is the Riemann tensor. The covariant derivative $\nabla_\mu$ is with respect to the Levi-Civita connection $\Gamma^{\mu}_{\ \nu \rho}$. The dilaton field is given by $\Phi$. The last term in the energy-momentum tensor $T$ suggests that the dilaton gives the Fradkin-Tseytlin term in the worldsheet
\begin{align}
\frac{1}{8\pi} \int d^2z \sqrt{\gamma} R_{\Sigma}\log(e^{-2\Phi}\sqrt{g}) 
\end{align}
where $R_{\Sigma}$ is the curvature of the worldsheet. The worldsheet metric $\gamma_{ab}$ and its determinant $\gamma$ should not be confused with the worldsheet superghost.

In order for them to be consistently gauged, these currents must be anomaly free. The anomaly-free condition is equivalent to the supergravity equations of motion (with $B_{\mu\nu} = 0$) in ten dimensions:
\begin{align}
R + 4 \nabla_\mu \nabla^\mu \Phi - 4 \nabla_\mu \Phi \nabla^\mu \Phi &= 0 \cr 
R_{\mu\nu} + 2\nabla_\mu \nabla_\nu \Phi & = 0  \ .
\end{align}
The equations of motion coincide with the supergravity limit (i.e. the leading order in $\alpha'$ expansions) of the superstring theory in ten dimensions.

In superstring theory, when we introduce a linear-dilaton background, the matter central charge is shifted so that we  may consider non-critical string theory (i.e. string theory not in ten dimensions). It turns out, however, in ambitwistor string theory, there is no singular $X\cdot X$ operator product expansion, and the central charge is not shifted. Thus, the non-critical ambitwistor string theory from the linear dilaton background is not available. We assume that the dimension of the target space is ten.

\section{Scattering amplitudes in light-like linear dilaton background}
From now on, we focus on a particularly simple solution of supergravity equations of motion given by the light-like linear dilaton background:
\begin{align}
g_{\mu\nu} &= \eta_{\mu\nu} \cr
\Phi &= v(X^0 - X^1) = V_\mu X^\mu \ . 
\end{align}
Hereafter, indices are raised and lowered by the ten-dimensional Minkowski metric $\eta_{\mu\nu} = \mathrm{diag}(-1,1,1,\cdots)$, and we sometimes use the Lorentz invariant inner product $A\cdot B = A_\mu B^\mu$.
Substituting the background into \eqref{currents}, we obtain
\begin{align}
   \mathcal{G} &= \psi^\mu \Pi_\mu - 2 \partial \psi^\mu V_\mu \cr
   \bar{\mathcal{G}} & = \eta^{\mu\nu} \bar{\psi}_\mu \Pi_\nu + 2\eta^{\mu\nu} \partial \bar{\psi}_\mu V_\nu  \cr
   \mathcal{H} &= \eta^{\mu\nu} \Pi_\mu \Pi_\nu - 2 \eta^{\mu\nu} \partial \Pi_\mu V_\nu \cr
   T &= \Pi_\mu \partial X^\mu - \frac{1}{2}(\bar{\psi}_\mu \partial \psi^\mu + \psi^\mu \partial \bar{\psi}_\mu) + V_\mu \partial^2 X^\mu \ . 
\end{align}
Note that $\mathcal{G},\bar{\mathcal{G}}, \mathcal{H}$ are all primary operators with respect to the energy-momentum tensor $T$. The operator product expansions among them suggest they are anomaly-free in ten dimensions once ghost contributions are added to $T$ so that the total central charge vanishes.

We should note that the {\it light-like} linear dilaton background is an exact solution both in superstring theory and in ambitwistor string theory. Because the dilaton gradient is light-like, the central charge is not shifted in superstring theory. If the dilaton gradient were not light-like, the superstring theory and the ambitwistor string theory would not share the same background.

In order to compute graviton scattering amplitudes in ambitwistor string theory, we need to introduce three types of vertex operators. Let us begin with the fixed-position vertex operator with picture number $-1$ \cite{Adamo:2017sze}\cite{Adamo:2018ege}.
\begin{align}
V_h =   c\bar{c} \delta(\gamma)\delta(\bar{\gamma}) O_h =  c\bar{c} \delta(\gamma)\delta(\bar{\gamma}) h^\mu_\nu \bar{\psi}_\mu \psi^\nu \ . 
\end{align}
We require that $V_h$ is BRST closed with respect to the BRST charge
\begin{align}
 Q_B = \oint dz (cT + bc\partial c + \frac{\tilde{c}}{2} \mathcal{H} + \bar{\gamma} \mathcal{G} + \gamma \bar{\mathcal{G}} + 2 \gamma \bar{\gamma} {\tilde{b}} ) \ .
\end{align}
(Note that $c,\tilde{c},b,{\tilde{b}},\gamma,\bar{\gamma}$ are all left-moving.)
It gives the gauge condition and the on-shell condition
\begin{align}
\partial^\mu h_{\mu\nu} &= 2h_{\mu\nu} V^\mu \cr
\partial^2 h_{\mu\nu} &= 2 \partial_\sigma h_{\mu\nu} V^\sigma \ .
 \end{align}

Let us choose the plane-wave basis: $h_{\mu\nu} = \epsilon_{\mu\nu} e^{ikX}$.
The gauge condition and on-shell condition become
\begin{align}
ik^\mu \epsilon_{\mu\nu} &= 2\epsilon_{\mu\nu} V^\mu \cr
-k^2 &= 2i k_\mu V^\mu \ . 
 \end{align}
Note that $V^2 = 0$ since $V_\mu$ is light-like.

The fixed-position vertex operator with picture number $0$ can be computed by a simple poles appearing in $\bar{\mathcal{G}} (\mathcal{G}O_h) - \mathcal{G}(\bar{\mathcal{G}} O_h)$ \cite{Adamo:2017sze}\cite{Adamo:2018ege}. The explicit computation gives
\begin{align}
U_h= \Pi_\mu \Pi_\sigma h^{\mu\sigma} - \Pi_\mu \bar{\psi}_\rho \psi^\sigma \partial^\rho h_{\sigma}^\mu + \Pi_\sigma \bar{\psi}_\mu \psi^\rho \partial_\rho h^{\mu\sigma} + \bar{\psi}_\rho \bar{\psi}_\mu \psi^{\sigma} \psi^{\lambda} \partial^\rho \partial_{\lambda} h^\mu_\sigma \ .     
\end{align}

Finally, the integrated vertex operator can be computed as
\begin{align}
\int d^2z \mathcal{V} = \int d^2z \bar{\delta}(\oint dz \mathcal{H}) U_h = \int d^2z \bar{\delta} ((k_\mu + 2iV_\mu)\Pi^\mu) U_h \ . 
\end{align}
See \cite{Ohmori:2015sha} for the first-principle derivation of the first equality.

With these ingredients, we can compute 
the tree-level amplitudes for $n$ gravition scattering in the ambitwistor string theory by evaluating the world-sheet path integral
\begin{align}
\mathcal{M}_n = \langle V_1 V_2 c_3 \tilde{c}_3 U_3 \int d^2z_4 \mathcal{V}_4 \cdots \int d^2z_n \mathcal{V}_n \rangle \ 
\end{align} 
on the sphere.
We can promote the $e^{ik_\mu X^\mu}$ factor in the vertex operator into the effective action
\begin{align}
S_{\mathrm{eff}} = \frac{1}{2\pi} \int d^2z \left( \Pi_\mu \bar{\partial} X^\mu + \bar{\psi}_\mu \bar{\partial} \psi^\mu + \frac{R_{\Sigma}}{2} V_\mu X^\mu - 2\pi i \sum_{i=1}^n k_{i\mu} X^\mu (z) \delta (z-z_i) \right) \ .
\end{align}

The path integral over $X^\mu$ can be done explicitly. The zero-mode integration gives the momentum (non-)conservation delta function
\begin{align}
\delta \left(\sum_{i=1}^n k_{i \mu} +2iV_\mu \right) \ .
\end{align}
Here, we have used the Gauss-Bonnet theorem $\frac{1}{4\pi} \int d^2z \sqrt{\gamma} R_\Sigma = 2$ on the sphere.

Before going on, let us make a small comment on the meaning of the delta function in the momentum (non-)conservation here. There are two possible interpretations. The first is to regard it as a diverging integral over a real line \cite{Hellerman:2008wp}
\begin{align}
\delta(k + iV) := \frac{1}{2\pi} \int_{-\infty}^{\infty} dx e^{ikx -Vx} \ .
\end{align}
Alternatively, we may define the delta function against the holomorphic variable $k+iV$ by modifying the contour of the $x$ integration. We will use the latter interpretation, which becomes useful at various places in our discussions below (e.g. to show the cyclic invariance of the scattering amplitudes). In relation, we assume that the momentum $k_\mu$  takes complex values.

The non-zero mode path integral of $X^\mu$ gives a delta functional constraint on $\Pi_\mu$:
\begin{align}
\bar{\partial} \Pi_\mu = -2\pi i \sum_i k_{i\mu} \delta^2 (z-z_i) + \frac{R_{\Sigma}}{2} V_\mu \ . \label{pieq}
\end{align}
Assuming that the curvature is localized at $z_* \to \infty$, we can integrate \eqref{pieq} to obtain
\begin{align}
\Pi_\mu = \sum_i\frac{-i k_{i\mu}}{z-z_i} + \frac{2V_\mu}{z-z_*} \ . \label{piin}
\end{align}
Here, we have used the formula $\bar{\partial} \frac{1}{z-w} = 2\pi\delta^2 (z-w)$. The integration constant is fixed by the requirement that $\Pi_\mu$ is a meromorphic one-form on the sphere.
We will eventually set $z_* = \infty$ and neglect the last term of \eqref{piin}. This will be the case even for the integrated vertex operators because the scattering equation will fix their location on the sphere and they do not generically approach $z_*$ as we will see.\footnote{{This choice of $z_*$ breaks the $\mathrm{SL}(2,\mathbb{C})$ invariance of the worldsheet, but as usual in the linear dilaton background in string theory, it only affects the zero-mode and the evaluation of the non-zero mode path integral can be done as if it were $\mathrm{SL}(2,\mathbb{C})$ invariant.}}

Let us begin with the simplest example of three-graviton scattering. The direct evaluation of the path integral over $\psi_\mu$, $\bar{\psi}^\nu$ gives
\begin{align}
\mathcal{M}_3 = -\frac{1}{4} \delta(\sum_{i=1}^3 k_{i \mu} + 2iV_\mu) \left( \epsilon_1 \cdot \epsilon_2 \epsilon_{3 \mu\nu} k_{12}^\mu k_{12}^\nu + 2\epsilon_{1\nu}^\mu \epsilon^\rho_{2\sigma} \epsilon^\sigma_{3\mu} k^\nu_{23 } k_{31 \rho} + \text{cyclic}  \right)   \ ,
\end{align}
where $k^\mu_{ij} = k^\mu_i -k^\mu_j$.
In order to make the expression manifestly cyclic invariant, we have used the gauge condition and the momentum (non-)conservation. For instance, we have used the identity
\begin{align}
k_{3 \mu} \epsilon^{\mu}_{1\nu} = \frac{k_{32 \mu} \epsilon^\mu_{1\nu}}{2} \ ,
\end{align}
which is true even when $V_\mu \neq 0$.

For more than four graviton scattering, we need to use the integrated vertex operators.
The delta function in the integrated vertex operators gives the scattering equation
\begin{align}
0 = \sum_{j\neq i} \frac{(k_i + 2i V) \cdot k_j}{z_i-z_j} \  
\label{sca} \end{align} 
(in the coordinate of the sphere where the curvature is localized at infinity).
Note that the summation at $j=i$ is removed because of the on-shell condition in the light-like linear dilaton background. The integration over $z_i$ will be localized at points specified by the scattering equation.

Let us note that the scattering equation of this form is $\mathrm{SL}(2,\mathbb{C})$ invariant only after taking into account the curvature localized at infinity. Translation and dilatation is obvious, so we will only check $z \to z^{-1}$. The right hand side of \eqref{sca} together with the contribution from $z_* = \infty$ transforms as
\begin{align}
&\sum_{j\neq i} \frac{z_i z_j (k_i + 2iV) \cdot k_j}{z_i-z_j} +\frac{z_i z_* (k_i + 2iV) \cdot 2iV}{z_i-z_*} \cr
=& \sum_{j\neq i} \frac{(z_i^2 - z_i(z_i-z_j)) (k_i + 2iV) \cdot k_j}{z_i-z_j} + \frac{(z_i^2 - z_i(z_i-z_*)) (k_i + 2iV) \cdot (2iV) }{z_i-z_*} \cr  
=& z_i^2 \left(\sum_{j\neq i} \frac{(k_i + 2iV) \cdot k_j}{z_i-z_j}+ \frac{(k_i + 2iV) \cdot (2iV)}{z_i-z_*} \right)  - z_i (k_i + 2iV) \cdot \left(\sum_{ j \neq i}  k_j +2iV \right) \ . \label{sc2}
\end{align}
At this point, one may use the (modified) momentum conservation $-2iV = k_i + \sum_{j \neq i} k_j $ and the on-shell condition $-k^2 = 2iV \cdot k$  to see that the last term in \eqref{sc2} vanishes, and the scattering equation is $\mathrm{SL}(2,\mathbb{C})$ invariant (or covariant with the contribution from the curvature at infinity properly transformed).

Following \cite{Mason:2013sva}, we can obtain the formal expression for the $n$-graviton scattering amplitudes:
\begin{align}
\mathcal{M}_n = \delta(\sum k_{i\mu} + 2iV_\mu) \int \frac{1}{\mathrm{Vol} \mathrm{SL}(2,\mathbb{C})} \mathrm{Pf}'(M_1) \mathrm{Pf}'(M_2) \prod_{i=4}^n \bar{\delta} ((k_i + 2iV) \cdot \Pi (z_i))  \ .
\end{align}
Here $\frac{1}{\mathrm{Vol} \mathrm{SL}(2,\mathbb{C})}  = \frac{z_{12}z_{23}z_{31}}{dz_1 dz_2 dz_3}$ comes from the ghost path integral (in the coordinate where the curvature is localized at infinity).\footnote{As we have mentioned in footnote 1, the choice of $z_*$ seems to break the $\mathrm{SL}(2,\mathbb{C})$ invariance, but since the relevant factor that cancels this ghost path integral is the non-zero mode path integral of fermions, which is $\mathrm{SL}(2,\mathbb{C})$ invariant, this nomenclature is reasonable. The direct computation of three-point graviton scattering indeed suggests that this factor makes the amplitudes independent of the position of 
 fixed-point vertex operators and cyclic invariant. 
 }
To explain the notation further, we first decompose the polarization tensor by $\epsilon_{\mu\nu} = \epsilon_\mu^1 \epsilon_{\nu}^2$. To compute the graviton scattering amplitudes, we take the symmetric part of the polarization tensor $\epsilon_\mu^1 \epsilon_{\nu}^2$ at the end of the computation. The formula actually computes the scattering of Kalb-Ramond field as well by considering the anti-symmetric part of $\epsilon_\mu^1 \epsilon_{\nu}^2$.

Here $M_1$ and $M_2$ are $2n$ by $2n$ matrices defined by 
\begin{align} M = 
\begin{pmatrix}
  A &-C^T  \\
  C & B \\ 
\end{pmatrix} \ ,
\end{align}
where each component is 
\begin{align}
A_{ij} = \frac{k_i \cdot k_j}{z_{ij}} \ , \ \ B_{ij} = \frac{\epsilon_i \cdot \epsilon_j}{z_{ij}} \ , \ \ C'_{ij} = \frac{i \epsilon_i \cdot k_j}{z_{ij}} \ 
\end{align}
Here, for $M_1$ we use the polarization vector $\epsilon^1_{i \mu}$ and for $M_2$ we use $\epsilon^2_{i\mu}$. 
To obtain $C$ (rather than $C'$), we have to replace the diagonal component of $C'$ with
\begin{align}
C_{ii} = - \sum_{j\neq i} \frac{i\epsilon_i \cdot k_j}{z_{ij}} \ 
\end{align}
while keeping the off-diagonal components the same as $C'$.

The final expression involves $\mathrm{Pf}'$, which is obtained by removing the first two rows and columns from $M$ and multiplying the ghost contribution 
\begin{align}
\mathrm{Pf}' M = \frac{1}{z_{12}}\mathrm{Pf}M_{12}^{12} \ . 
\end{align}
This is originated from treating two vertex operators (i.e. $V_1$ and $V_2$ here) in a special manner, but the final results should not depend on the choice as we explicitly saw in the three-graviton scattering.

Since the path integral over fermions does not depend on $V^\mu$, the apparent difference from the Minknowski space-time is only in the momentum (non-)conservation. Of course, because the on-shell condition and the gauge condition are different, the actual scattering amplitudes do depend on $V^\mu$ non-trivially.

\section{Discussion}

Let us compare our tree-level amplitudes with the scattering amplitudes of the (super)string theory in the light-like linear dilaton background. As for the three-particle scattering, our results agree with the one in the literature \cite{Ho:2007ar}\cite{Chan:2009eh} in the sense that the only modification is momentum conservation.\footnote{This is true only in type II superstring theory and type II supergravity. In heterotic superstring theory and in supergravity, the three graviton scattering amplitudes differ by $\alpha'$ corrections.} For the more than four-particle scattering, there should be a difference because the (super)string theory admits an infinite number of poles due to the exchange of the stringy tower, but the ambitwistor string theory does not have these states. Accordingly, the high-energy behavior must be different.

How about the quantum corrections?
The ambitwistor string theory may be formally defined on the higher genus worldsheet although the amplitudes can be ultraviolet divergent unlike in the superstring theory. At the formal level, we imagine that the momentum (non-)conservation in genus $g$ amplitudes becomes
\begin{align}
\delta \left(i \sum_i k_\mu^i + 2(g-1)V_\mu \right) \label{genusg}
\end{align}
from the zero mode path itnegral over $X$ by using the Gauss-Bonnet theorem $\frac{1}{4\pi} \int d^2z \sqrt{\gamma} R_{\Sigma} = -2(g-1)$. This selection rule gives an interesting consequence: each scattering amplitude with fixed eternal momenta has contributions only from a particular genus without further corrections. 

The selection rule can be understood from the field theory language as follows. In the light-like linear dilaton background, the momentum space propagator has a delta function $\delta(k_1+k_2 -2iV)$ while the vertex has a delta function $\delta(\sum k +2iV)$ (in the convention that the incoming line at each vertex has a plus sign for the momentum). This is because the tree-level action has the schematic form of $\int d^{10} x e^{-2V x}(-\partial_\mu h \partial^\mu h -H_{\mathrm{int}}(h))$. Consider the tree-level $n$ graviton scattering amplitudes constructed out of three-point vertices. It has $n-2$ vertices with $n-3$ propagators, leading to the overall factor of $\delta(\sum_i^n k_i +2iV)$. {One can generalize the story in loop amplitudes with genus (or loop number) $g$. Assuming that the amplitudes are constructed out of $n-2+2g$ three-point vertices with $n-3+3g$ propagators, we have precisely the conservation law of \eqref{genusg}.}

{This field-theoretic interpretation of the selection rule helps us understand the factorization property of the scattering amplitudes. For illustration, we focus on the limit $z_1 \to z_2$ in the scattering equation. In this limit, the scattering equation demands $(k_1+ 2iV) \cdot k_2 = (k_2+2iV)\cdot k_1 = 0$. This is equivalent to the condition that $-(k_1+k_2 + 2iV)^2 = 2i (k_1+k_2 + 2iV) \cdot V$, which means that the internal propagator is on-shell in accordance with the rule above. In this way, we see that the degeneration of the worldsheet corresponds to the internal on-shell propagator and the factorization of the amplitudes. While we have demonstrated it in the simplest example, the degeneration of the scattering equations with a more complicated limit can be studied \cite{Ohmori:2015sha}\cite{Geyer:2015jch}.}
Another potentially singular limit is when $z_i$ approaches the curvature at infinity $z_*=\infty$, which demands $k_i \cdot V = k_i^2 = 0$. It corresponds to the fact that one can freely produce particles with momentum proportional to $V$ in this background. If a particle has $k_i \cdot V = k_i^2 = 0$, then the production of an (on-shell) particle with momentum $2i\alpha V_i$ keeps it still on-shell for arbitrary $\alpha$.

Finally, the linear dilaton theory has a strong coupling problem. A typical way to resolve the strong coupling problem in the bosonic string theory is to introduce the Liouville potential in the worldsheet. A similar technique may be used in superstring theory or ambitwistor string theory by possibly replacing tachyon with graviton. The study of such background is much more complicated and further investigation is necessary.

\section*{Acknowledgements}
This work by YN is in part supported by JSPS KAKENHI Grant Number 21K03581. YN would like to thank P.~M.~Ho for a small chat on the meaning of the delta function in the momentum (non-)conservation.

\end{document}